\documentclass[reprint,aip,jcp,floatfix,unsortedaddress]{revtex4-1}
\usepackage{graphicx}
\usepackage{dcolumn}
\usepackage{longtable}
\usepackage{multirow}
\usepackage{textcomp}
\usepackage{amsmath}
\usepackage{rotating}
\usepackage{tikz}
\usepackage{siunitx}
\usepackage[modulo]{lineno}

\newcommand{\ie}{i.e.\ }
\newcommand{\eg}{e.g.\ }
\newcommand{\etal}{{\em et al.}\ }

\newcommand{\amol}{\alpha_\mathrm{mol}}
\newcommand{\Vmol}{V_\mathrm{mol}}

\newcommand{\NA}{N_{\!\mathrm{A}}}
\newcommand{\nD}{n_\mathrm{D}}
\newcommand{\mol}{\mathrm{mol}}
\newcommand{\aIL}{\alpha_\mol}
\newcommand{\vaIL}{\vec{\alpha}_\mol}
\newcommand{\aT}{\alpha_\mathrm{types}}
\newcommand{\vaT}{\vec{\alpha}_\mathrm{types}}
\newcommand{\VIL}{V_\mol}

\begin{document}
\title{Additive polarizabilities in ionic liquids}
\author{Carlos ES Bernardes, Karina Shimizu and Jos\'e Nuno Canongia Lopes}
\affiliation{Centro de Quimica Estrutural, Instituto Superior T\'ecnico, Universidade de Lisboa, Portugal}
\author{Philipp Marquetand}
\affiliation{University of Vienna, Institute of Theoretical Chemistry, Austria}
\author{Esther Heid, Othmar Steinhauser, Christian Schr\"oder}
\email{christian.schroeder@univie.ac.at}
\affiliation{University of Vienna, Department of Computational Biological Chemistry, Austria}
\date{\today}

\begin{abstract}
An extended Designed regression analysis of experimental data on density and refractive indices of several
classes of ionic liquids yielded statistically averaged atomic volumes and polarizabilities of the constituting atoms.
These values can be used to predict the molecular volume and polarizability of an unknown ionic liquid as well as its 
mass density and refractive index. Our approach does not need information on the molecular structure of the ionic 
liquid, but it turned out that the discrimination of the hybridization state of the carbons improved the overall result.
Our results are not only compared to experimental data but also to quantum-chemical calculations. Furthermore,
fractional charges of ionic liquid ions and their relation to polarizability are discussed.
\end{abstract}
\maketitle
\clearpage{}

\section{Introduction}
The classical MD force field\cite{pad04a,pad12a} of Canongia Lopes and P{\'a}dua has been quite successful to reproduce structural properties of ionic liquids, but unfortunately
the simulated systems were too viscous due to very strong directional Coulomb forces between the ions.\cite{mag07b,bor09a} A cheap but 
dangerous workaround is the scaling of all partial charges by a uniform factor. 
This accelerates the dynamics without any additional computational effort but destroys almost all hydrogen bonds in the ionic liquids and
may lead to spurious artifacts, \eg  for molecular rotations.\cite{sch12a} Furthermore, the scaling factor has to be determined for each 
combination of cation and anion separately. Consequently, when applying this scaling factor for a particular cation-anion combination,
partial charges of the cation are not transferable anymore, \ie cationic partial charges differ in a simulation 
of 1-butyl-3-methylimidazolium chloride [C$_4$C$_1$im]Cl from those of 1-butyl-3-methylimidazolium bis(trifluoromethylsulfonyl)imide [C$_4$C$_1$im][NTf$_2$].

Since we are interested in a general, transferable force field for ionic liquids to reproduce experimental data of ionic liquids 
and in particular their dynamics, we aim for polarizable forces in addition to the classical force field. Borodin \etal already published
a polarizable force field in 2009,\cite{bor09a} but the access of parameter is restricted and Buckingham potentials are used 
instead of more common Lennard-Jones potentials. Furthermore, many of these intermolecular interactions are given pairwise and not on
the basis of mixing rules. This makes the addition of new compounds quite tedious and as a result the force field less transferable.
Here, we would like to present the first step to update the successful Canongia Lopes-P\'adua force field by including polarizable forces.
Several authors have shown that the impact of polarizability on the structure is of minor importance.\cite{vot04b,ste10b}
Therefore, we prefer to keep the existing force field parametrization as much as possible and include polarizable forces to accelerate dynamics.

In principle, several ways exist to implement polarizable forces -- 
the most prominent methods are Drude oscillators and induced-point dipoles,\cite{gun05a}
which coincide quite well for ionic liquids.\cite{ste15a} In contrast to the fluctuating charge model\cite{gun05a},
Drude oscillators and induced-point dipoles require only the atomic polarizabilities as additional force field parameters.
Molecular polarizabilities $\aIL$ are easily accessible via quantum-chemical calculations or via the Lorentz-Lorenz equation from experimental
refractive indices $\nD$ and molecular volume data, $\VIL$.
\begin{equation}
 \frac{\nD^2-1}{\nD^2+2} = \frac{4 \pi}{3} \frac{\aIL}{\VIL}
 \label{EQ_LL}
\end{equation}

In principle, several ways to determine atomic polarizabilities exist: In addition to empirical models,\cite{sav79a} Designed regression\cite{sed13a} and 
similar statistical approaches may predict the polarizability on the basis of the molecular topology (but unknown atomic coordinates) assuming that
the polarizabilities of neighboring atoms do not interact. More sophisticated methods use the atomic coordinates to compute 
averaged polarizabilities for important chemical elements including the impact of vicinal atoms via the dipole-dipole tensor.\cite{swa98a,jah06a,tur11a,yan13a}
Polarizabilities are also accessible via the distributed polarizability model\cite{wip14a} or 
general automatic atomic model parametrization (GAAMP).\cite{rou13a}

Although many refractive indices and mass densities of imidazolium-based ionic liquids could be reproduced by uniform polarizabilities for 
each chemical element,\cite{sed13a} our model failed for ionic liquids containing dicyanamides. Most probably this is due to the fact that
the polarizability of the sp-nitrogens and carbons of the anion are not well represented by the statistical averaged values of nitrogen
and carbon mainly derived from corresponding sp$^3$ and sp$^2$ atoms in the imidazolium ring and the aliphatic chain. 
This inspired us to take into account the hybridization of the carbon atoms. The decomposition of 
molecular polarizabilities $\aIL$ and molecular volumes $\VIL$ into atomic contributions will be performed with Designed Regression 
assuming no interaction between the atomic polarizabilities.

\section{Designed Regression}
The molecular polarizabilities $\aIL$ of ionic liquid pairs can be determined in a two step process:
Starting from the experimental mass density $\rho$ of the compound, the molecular volume $\VIL$ can be determined by
$ \VIL = M \rho^{-1}\NA^{-1}$
using the molar mass $M$ and the Avogadro constant $\NA$. These molecular volumes $\VIL$ can be decomposed into atomic 
contributions, $V_\mathrm{types}$,  by classical one-way analysis of variance (ANOVA) method,\cite{sea87a,job91a,age00a,min09a} 
which is called  Designed Regression in MATHEMATICA:\cite{mathematica}
\begin{equation}
\vec{V}_\mol = \pmb{X} \cdot \vec{V}_\mathrm{types}
\end{equation}
The design matrix $\pmb{X}$ correlates a vector $\VIL$ containing molecular volumes of all ionic liquids under investigation
with a vector $\vec{V}_\mathrm{types} =$\{ $V_\mathrm{H}$, $V_\mathrm{B}$, $V_\mathrm{Csp^3}$, $V_\mathrm{Csp^2}$, $V_\mathrm{Csp}$,
$V_\mathrm{N}$, $V_\mathrm{O}$, $V_\mathrm{F}$, $V_\mathrm{P}$, $V_\mathrm{S}$, $V_\mathrm{Cl}$\} containing the statistically averaged contributions of each atom type.
Each row of $\VIL$ and the design matrix $\mathbf{X}$ contains information on a particular ionic liquid under investigation.
Each column of $\mathbf{X}$ contains the number of atoms of a specific atom type which are present in these ionic liquids.
In principle, it is possible to include several molecular volumes of the very same ionic liquid, \eg from different experimental 
measurements of the mass density, to minimize the error from impurities. Of course, the respective rows of $\mathbf{X}$ are identical
making the respective equations linearly dependent. However, this can be used to weight the importance of various experimental data on the very 
same ionic liquid for the statistical analysis. More information of $\mathbf{X}$ are given in the supplementary material.

In a second step, the molecular volume $\VIL$ and the refractive index $\nD$ are used to compute the molecular polarizability $\aIL$ of an ion pair 
via the Lorentz-Lorenz equation~\eqref{EQ_LL}. 
These $\aIL$-values are also subject to a Designed Regression analysis again assuming that these atomic contributions do not correlate with each other:\cite{sed13a}
\begin{equation}
\vaIL = \pmb{X} \cdot \vaT
\end{equation}

Using the experimental data cited in Ref.~\onlinecite{sed13a} and from IL Thermo\cite{ilthermo} (837 experimental mass densities and 414 experimental $\nD$ in total) 
we derive the values listed in Table~\ref{TAB_DR} for the polarizabilities of the respective atom types, $\vaT$, and their contribution to the molecular volume, $\vec{V}_\mathrm{types}$, 
of the ionic liquid ion pair.
\begin{table}[t]
 \caption{Atomic polarizabilities and volumes gained from the Designed Regression analysis in this work\label{TAB_DR}. Standard errors are given in the supplementary material.}
\begin{ruledtabular}
 \begin{tabular}{lrrrrrr}
           & \multicolumn{2}{c}{this work} & Ref.~\onlinecite{sed13a} &Ref.~\onlinecite{jho82a} & Ref.~\onlinecite{mil90b} & Ref.~\onlinecite{yan13a}\\
  Atom &  $V_\mathrm{types}$ & $\aT$ & $\aT$ & $\aT$ & $\aT$& $\aT$\\
            & [{\AA}$^3$] &  [{\AA}$^3$] & [{\AA}$^3$] & [{\AA}$^3$] & [{\AA}$^3$]& [{\AA}$^3$]\\
  \hline 
  H          &  5.91 & 0.389 & 0.210 & 0.386 & 0.387 & 0.444\\
  B          & 18.15 & 0.243 & 1.186 &       &       & 1.085\\
  C (sp$^3$) & 15.84 & 1.081 & 1.368 &1.064 & 1.061 & 1.152\\
  C (sp$^2$) & 15.27 & 1.290 & 1.368 &1.382 & 1.352 & 1.152\\
  C (sp)     & 20.04 & 1.292 & 1.368 &      & 1.283 & 1.152 \\
  N          & 14.35 & 1.085 & 1.123 &1.094 & 0.964 & 0.917\\
  O          &  9.28 & 0.354 & 0.641 & 0.664 & 0.637  & 0.331\\
  F          & 13.26 & 0.346 & 0.400 &     & 0.296  & 0.256\\
  P          & 22.99 & 1.098 & 1.768 &1.743 & 1.538 & 1.750\\
  S          & 37.28 & 2.771 & 2.376 &        & 2.700 & 2.703\\
  Cl         & 26.08 & 2.428 & 3.451&        & 2.315 & 2.138\\
 \end{tabular}
 \end{ruledtabular}
\end{table}
Originally, we also took the hybridization of nitrogens, oxygens and sulfurs into account but it turned out that the predictions did not improve.
Moreover, the assignment of sp$^3$, sp$^2$ and sp is far from easy in these compounds.\cite{bro11a,smi90a} For example, the central nitrogens
in N(CN)$_2^-$ and NTf$_2^-$ are sp$^2$ hybridized. The best criterion to decide the state of hybridization seems to be the angle. Below a 
threshold of 115$^\circ$ the central atom of that angle should be sp$^3$. However, this analysis necessitates coordinates and is not
applicable for sulfur in SCN$^-$ for example. Since carbons are usually not terminal atoms and possess no free electron pairs, the determination
on the basis of single, double and triple bonds is much easier. 
Furthermore, the hybridization state of the carbon has a significant influence on the polarizability as expected from our former study in case 
of dicyanimides.\cite{sed13a} The new values presented in this work shows a remarkable improvement in the prediction of these compounds
(Fig.~\ref{FIG_nd}). The different polarizabilities of sp$^2$ and sp$^3$ carbons were also reported by Kang \etal\cite{jho82a}
which agree reasonably well with ours. 
The hybridization of the carbon in the present work also affects the polarizability of the hydrogens. The new values 
have increased compared to Ref.~\onlinecite{sed13a} and now resemble more those values of Ref.~\onlinecite{jho82a,mil90b,yan13a}.
Based on Ref.~\onlinecite{mon02a} the polarizability of $\alpha_H = $\SI{0.389}{\angstrom^3} corresponds to a confinement of its electron within
\SI{2.3}{\angstrom}, which seems reasonable for a bounded hydrogen.

In contrast to our former study,\cite{sed13a} values for boron, oxygen, phosphorus and chlorine changed, which is due to the inclusion 
of non-imidazolium based ionic liquids in the present study. However, $\alpha_\mathrm{O}$ and $\alpha_\mathrm{Cl}$ are now closer to values reported 
for covalently bound oxygens and chlorines,\cite{mil90b,yan13a} but lower than corresponding values for ions.\cite{jah06a,tur11a}
The importance of covalent bonds for the atomic polarizability is striking for fluorine. In ionic liquids, fluorines are almost exclusively part 
of polyatomic anions, \eg triflate and NTf$_2^-$. As a result, our $\alpha_F$ values are quite low compared to the polarizability of ``free''
fluorines in Ref.~\onlinecite{tur11a}. There, Molina \etal also showed that the compression of MgO crystalline phases resulted in a decrease of 
the oxide polarizability which also might indicate that the close proximity of other atoms, \eg due to a chemical bond, lowers the respective
polarizabilities of the bound atoms.

\section{Comparison to experimental data}
Based on the polarizabilities and volumes in Table~\ref{TAB_DR}, one may compute the molecular polarizability $\aIL$ and molecular volume $\VIL$
of an unknown ionic liquid. From these values, the refractive index $\nD$ and the mass density $\rho$ can be predicted. 
\begin{figure}[b]
 \includegraphics[width=\linewidth]{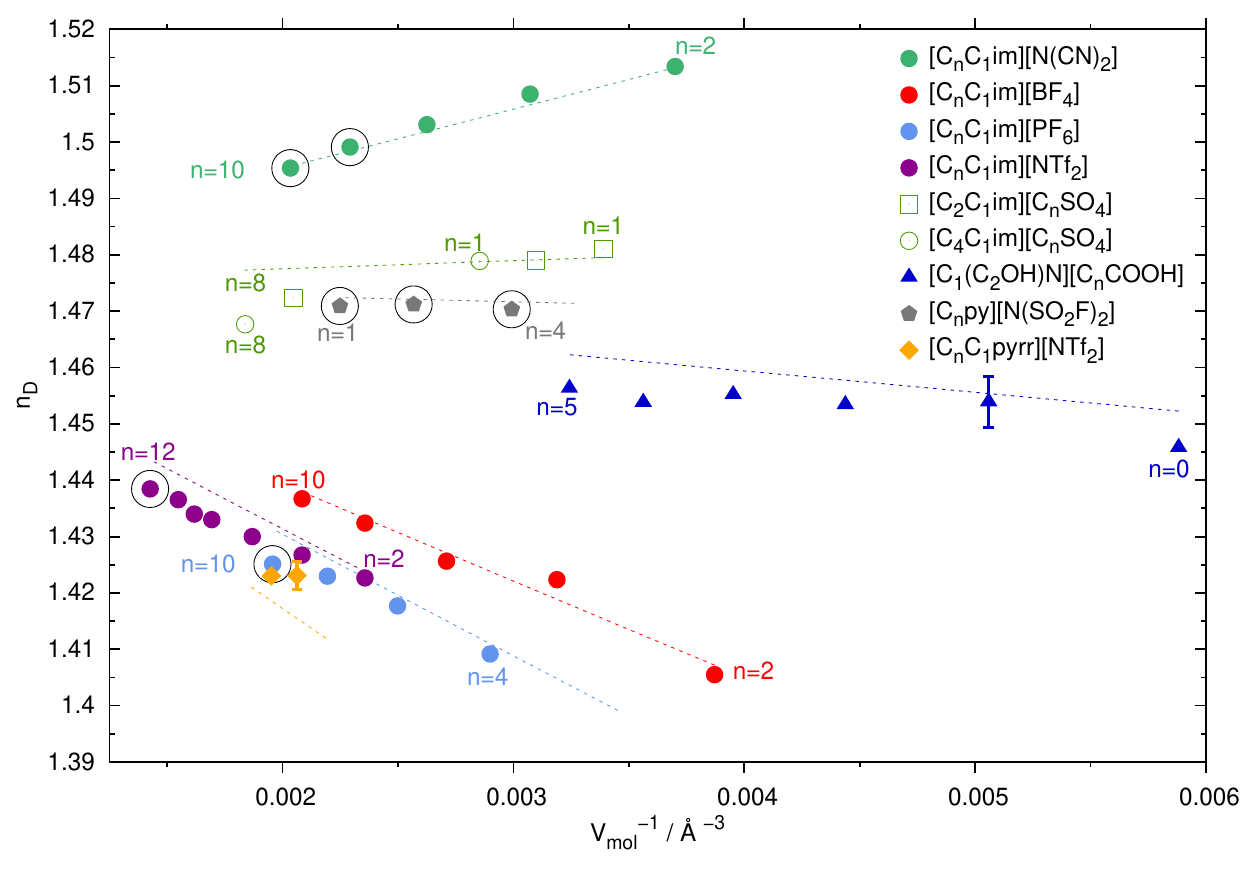}
 \caption{Comparison between experimental (points) and predicted (lines) refractive indices of various ionic liquids. 
 The circles indicate compounds which were not part of the learning set.\label{FIG_nd}}
\end{figure}
The agreement between predicted values (lines) and experimental data (points) is depicted in Fig.~\ref{FIG_nd} for homologous series of various cation/anion combinations.
Since the majority of our learning set consists of imidazolium-based ionic liquids, our prediction works best for them. In case of imidazolium alkylsulfates, both
the side chain of the cation and the aliphatic chain of the anion can be extended by CH$_2$-units, which results in the very same change of the refractive index within our model. 
This prediction seems valid as visible for the open green squares and circles in Fig.~\ref{FIG_nd}.
The biggest improvement compared to our former work\cite{sed13a} concerns imidazolium-based dicyanamides (turquoise circles) due to the discrimination between sp, sp$^2$ and sp$^3$
carbons. For example, 1-ethyl-3-methyl-imidazolium dicyanamide [C$_2$C$_1$im][N(CN)$_2$] has an experimental refractive index of 1.510. Using the uniform polarizabilities for each 
chemical element in Ref.~\onlinecite{sed13a} a refractive index of 1.537 is predicted. However, our new approach based on the values in Table~\ref{TAB_DR} predicts a $\nD$-value 
of 1.513.

\begin{figure}[b]
 \includegraphics[width=\linewidth]{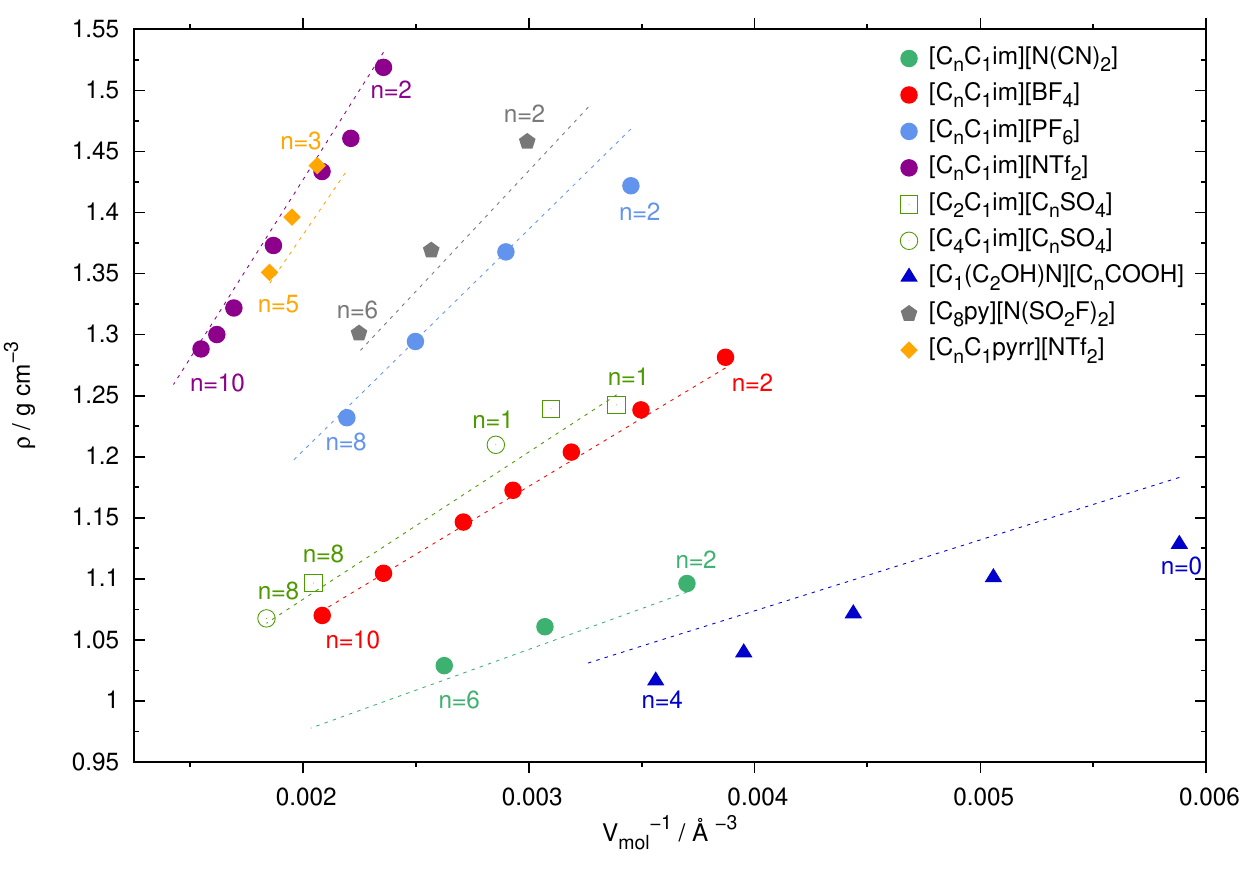}
 \caption{Comparison between experimental (points) and predicted (lines) mass densities of various ionic liquids.\label{FIG_rho}}
\end{figure}

Although not part of the learning set (as indicated by open black circles) computational $\nD$-values of pyridinium bis(fluorosulfonyl)amide [C$_n$py][N(SO$_2$F)$_2$] coincide quite well with corresponding experiments.
Unfortunately, the agreement is only reasonable (less than 1\% deviation) in case of the methyl-2-hydroxyethylammonium- [C$_1$(C$_2$OH)NH$_2^+$] and pyrrolidinium-based [C$_n$C$_1$pyrr$^+$] ionic liquids. 
This fact cannot be attributed to experimental uncertainties only, as visible for example for methyl,propyl-pyrrolidinium bis(trifluoromethylsulfonyl)imide  [C$_3$C$_1$pyrr][NTf$_2$] (orange squares) in Fig.~\ref{FIG_nd}.
In principle, this deviation can be due to non-optimal polarizabilities and/or volumes in our predictions (see Eq.~\ref{EQ_LL}). Fig.~\ref{FIG_rho} shows the experimental (symbols) and computational predictions (lines) 
of the mass density as a function of the reciprocal molecular volume revealing discrepancies between experimental and predicted values for ammonium and pyrrolidinium salts again.
Consequently, the source of error for the prediction of the refractive index is not the atomic polarizabilities in Table~\ref{TAB_DR} but the atomic volumes for those particular ionic liquids.
Nevertheless, the predicted densities can be used as a first guess to setup a simulation box. Afterwards, one should perform npT simulations to reach an equilibrium.

\section{Model of the excess electron}
The comparison to experiment is on the level of ion pairs, but not single molecular ions. However, the polarizabilities of the single molecules can be compared to quantum-chemical
calculations. Therefore, we performed geometry optimizations and frequency calculations using ORCA\cite{nee12a} on the level of MP2/aug-cc-pVDZ and CCSD(T)/aug-cc-pVQZ of several neutral, 
cationic and anionic species as shown in Table~I in the supplementary material. Both types of quantum-chemical calculation yield similar results indicating that the molecular polarizability
is well described already with MP2, also in the few cases where CCSD(T) was not feasible due to the molecular size.

\begin{figure}[b]
 \begin{tikzpicture}
   \node [] at (0.0,0.0) {\includegraphics[]{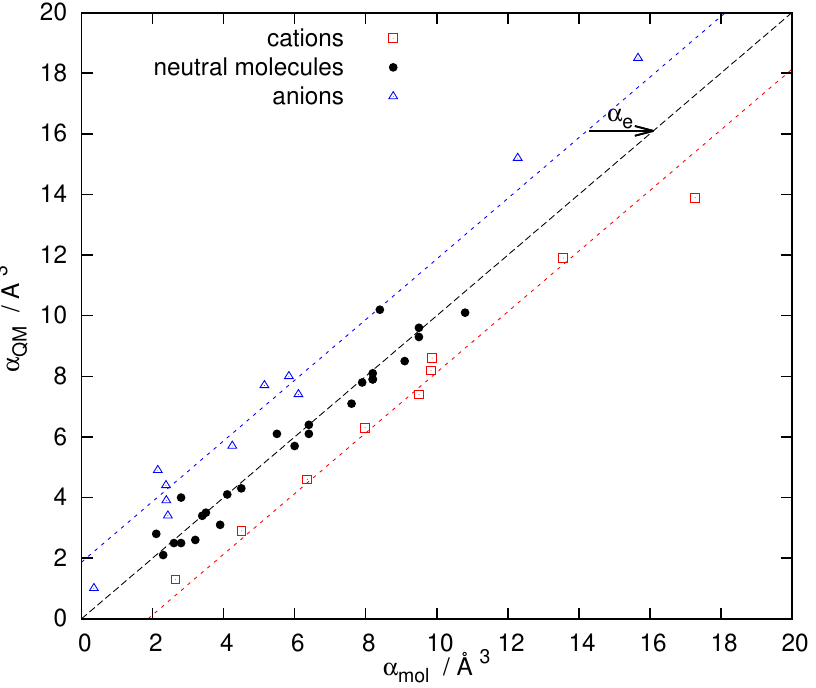}};
 \end{tikzpicture}
 \caption{Comparison between polarizabilities derived from MP2/aug-cc-pVDZ calculations and Designed Regression. The black dashed line represents
 100\% agreement. The blue and red lines correspond to an agreement when taking $\alpha_e =$~\SI{1.87}{\angstrom^3} into account. \label{FIG_aIL}}
\end{figure}
Comparing the results of the quantum-chemical calculations and the Designed Regression, one notices that the agreement for neutral 
compounds (black dots in Fig.~\ref{FIG_aIL}) is excellent. In contrast, the molecular polarizabilities of cations are overestimated by Designed 
Regression whereas the corresponding values for the anions are underestimated, which might be explained by our ``model of the excess electron''.
The net negative charge enables the anions to have more deformable electron clouds. Cations, however, have additional positive charge to hinder the electron
delocalization, which reduces the polarizabilities of the cation $\alpha^+$ and increases the polarizability of the anion $\alpha^-$.
In order to take this effect into account, we may model the ``polarizability of the excess electron $\alpha_e$''
\begin{eqnarray}
 \alpha^+ &=& \Bigl(\sum\limits_\beta \alpha_\beta \Bigl) - \alpha_{e} \\
 \alpha^- &=& \Bigl(\sum\limits_\beta \alpha_\beta \Bigl) + \alpha_{e}.
\end{eqnarray}

This excess polarizability $\alpha_e$ can be determined from the deviations of cations ($-\alpha_e=$~\SI{-1.82}{\angstrom^3}) and anions ($\alpha_e =$~\SI{1.92}{\angstrom^3})
between the MP2/aug-cc-pVDZ and the Designed Regression results separately.
However, in Fig.~\ref{FIG_aIL} we have used the average value of $\alpha_e =$~\SI{1.87}{\angstrom^3} (blue and red dashed line). 
Please note that the cations and anions investigated here are quite diverse (see Table~I supplementary material) and therefore the 
agreement of the shift is amazing. Unfortunately, it only holds true for monovalent ions. However, they represent the vast majority of ionic liquids. 
Furthermore, applying our model of the ``polarizability of the excess electron'' reproduces many polarizabilities of the ions in Table 1 of Ref.~\onlinecite{far09a}
and explains to some extent the higher polarizability of the oxide anions in Ref.~\onlinecite{jah06a}.
The excess polarizability can also be deduced from pure quantum-chemical calculations. For example, the molecular polarizability of CN$^-$ and SCN$^-$
is roughly \SI{1.7}{\angstrom^3} higher than those of HCN and HSCN, respectively. The additional proton decreases the mobility of the electrons resulting in a lower 
overall polarizability. A simple Designed Regression predicts the opposite due to the polarizability of the hydrogen. This fact shows the importance
of our $\alpha_e$-model, since in simulations the charge delocalization of each molecular ion matters:
The polarizability of the ``excess electron'' $\alpha_e$ in the anions or its depletion in the cations should be uniformly distributed among the corresponding atoms 
$\alpha_\beta^\mathrm{eff} = \alpha_\beta \pm \alpha_e/n$. 
As a result the effect of the excess polarizability on each atom $\beta$ decreases with increasing number of atoms $n$ of the ion.
Since the sum $\alpha^+ + \alpha^-$ equals the polarizability of the ion pair $\aIL$, our predictions of the ionic liquids in the last section 
are unaffected by this model.

\section{Molecular polarizability and partial charge scaling}
A cheap workaround often used to accelerate the dynamics of ionic liquids in computer simulation is the downscaling of partial charges which seems justified 
by the fact that net charges of the cation and the anion yields fractional values in quantum-chemical calculations of single ion pairs.
Of course, due to lower Coulombic interactions transport properties of charged scaled simulations are closer to experiment.\cite{sch12a,sal15a} This acceleration is a non-linear
function of the charge scaling factor and may lead to ion dynamics in simulations which are faster than experiment.\cite{tur09a,sch12a}

In principle, the fractional charges in the quantum-chemical calculations of the ion pair arise from the overlap of diffusive wavefunctions and 
subsequent assignment of the electron density to the cationic / anionic part based on the algorithm of the partial charge method, \eg ChelpG. 
Besides applying the charge scaling to the MD simulation, many authors interpret the scaling factor in a physical/chemical meaningful way. 
One may discuss the fractional charge in terms of a ``charge transfer'' from the anion towards the cation,\cite{ste10c} in particular for 
strongly coordinating anions.
Technically speaking, this interpretation may be correct but it implies that this charge transfer is of long-living nature and only occurs 
between exactly one anion and one cation. In ionic liquids, however, some cation-anion contact lasts only for a few picoseconds, others may live for several nanoseconds. 
Furthermore, a cation is surrounded by several anions, \eg eight in case of [C$_2$C$_1$im][OTf],\cite{sch11a} and many cations. This fact hampers 
the choice of a single ion pair in which the charge transfer takes place. It is even more complicated in mixtures of ionic liquids with other 
molecular solvents since the ions will not only be surrounded by counter-ions and ions of like charge but also by polar neutral molecules, \eg water.
All these circumstances make a discussion in terms of a charge transfer difficult.

The second way of interpretation is the polarizability of the system. The charge of the cations and anions is delocalized within the molecule and may 
react to approaching cations and anions. Here, the polarizable forces act like an ``inner solvent`` or dielectric continuum.\cite{mue07a,stu09a,sch12a}
In other words, the charge scaling takes the average polarizable forces into account. Indeed, we could show that polarizable forces in ionic liquids
reduce the Coulombic interaction.\cite{sch12a}
\begin{figure}[t]
 \begin{tikzpicture}
  \node [] at (0.0,0.0) {\includegraphics[width=\linewidth]{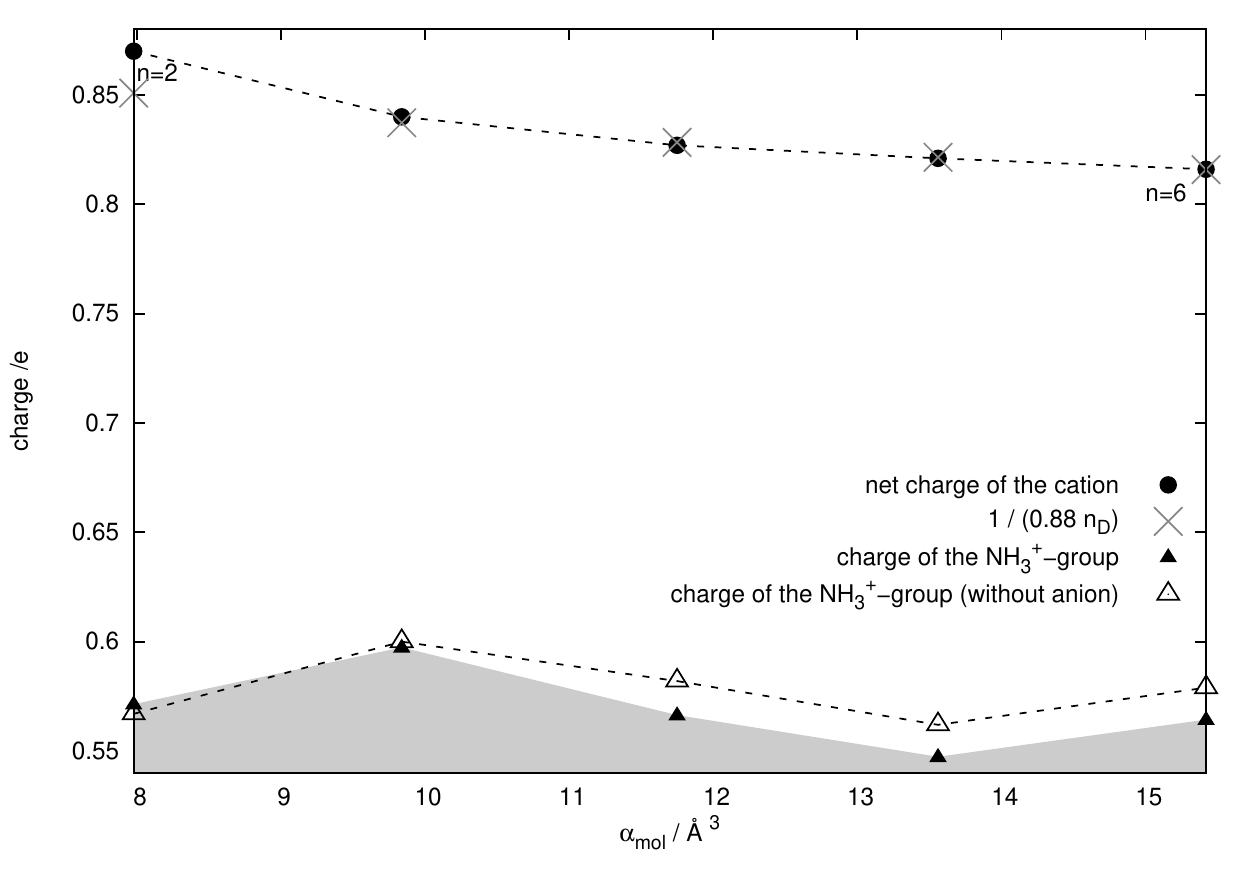}};
  \node [rotate=40] at (-1.2,0.7) {\includegraphics[width=3cm]{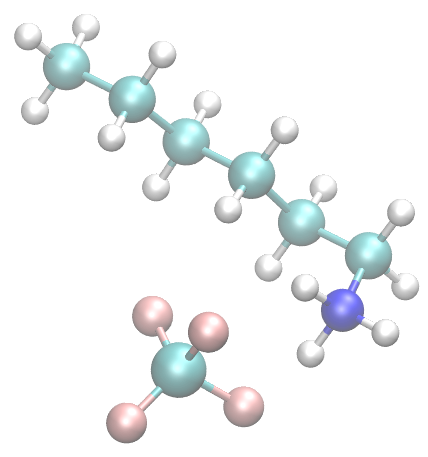}};
  \end{tikzpicture}	
 \caption{Charge scaling factor (= sum of all cationic partial charges) as a function of polarizability gained from a ChelpG analysis using MP2/aug-cc-pVDZ. 
 The inset represents the optimized geometry of [C$_6$NH$_3$][BF$_4$].\label{FIG_scaling} }
\end{figure}
In order to contribute further to this discussion, we optimized the geometry and subsequently calculated the partial charges $q_i$ of a homologous
series of alkylammonium tetrafluoroborates [C$_n$NH$_3$][BF$_4$] with $n = 2 \ldots 6$ using MP2/aug-cc-pVDZ and ChelpG. The charge of the cation and 
anion is
\begin{eqnarray*}
 q^+ &=& \sum\limits_i^\mathrm{cation} q_i = q^\mathrm{head}  + q^\mathrm{tail}  \le \mathrm{1 e} \\
 q^- &=&  \sum\limits_i^\mathrm{anion} q_i = - q^+
\end{eqnarray*}
The position of the anion (irrespective of the corresponding $n$) is close to the polar head group NH$_3^+$ as shown in the inset of Fig.~\ref{FIG_scaling} 
and expected by chemical intuition. The black circles represent $q^+$ (when calculated together with the anion) and also refer to the charge scaling factor $q^+ / \mathrm{1 e}$. 
With increasing chain length $n$, the charge scaling factor decreases. However, the net partial charge of the NH$_3^+$ ($q^\mathrm{head}$, filled triangles) does not follow this trend. 
Moreover, the net partial charges of the polar head group $q^\mathrm{head}$ changes only slightly, if no anion is present in the quantum-chemical calculation 
(open triangles in Fig.~\ref{FIG_scaling}), \ie when no charge transfer is possible.
The charge transfer should take place between the groups with the highest interaction but this is obviously not the case for the alkylammonium tetrafluoroborates
[C$_n$NH$_3$][BF$_4$]  since $q^\mathrm{tail}$ of the apolar side alkyl chain is affected much more by the presence of the anion than $q^\mathrm{head}$.

Since we study only one anion in the neighborhood of the alkylammonium the local field at the terminal carbons 
exerted by that anion becomes weaker explaining the leveling-off of the partial charge factor in Fig.~\ref{FIG_scaling}.
However, the polarizability of the cation increases with increasing chain length and therefore the possibility for charge delocalization. Dielectric continuum
theories\cite{mue07a,stu09a,sch12a} correlate the scaling factor with $1/\sqrt{\epsilon_\infty} = 1/ \nD$ using the high frequency limit of the dielectric 
constant $\epsilon_\infty$ which depends on the macroscopic polarizability of the sample. The prefactor of 0.88 in Fig.~\ref{FIG_scaling} is gained by a fit
and indicates that the dielectric screening by the polarizabilities is not completely effective.

\begin{figure}[b]
 \includegraphics[width=\linewidth]{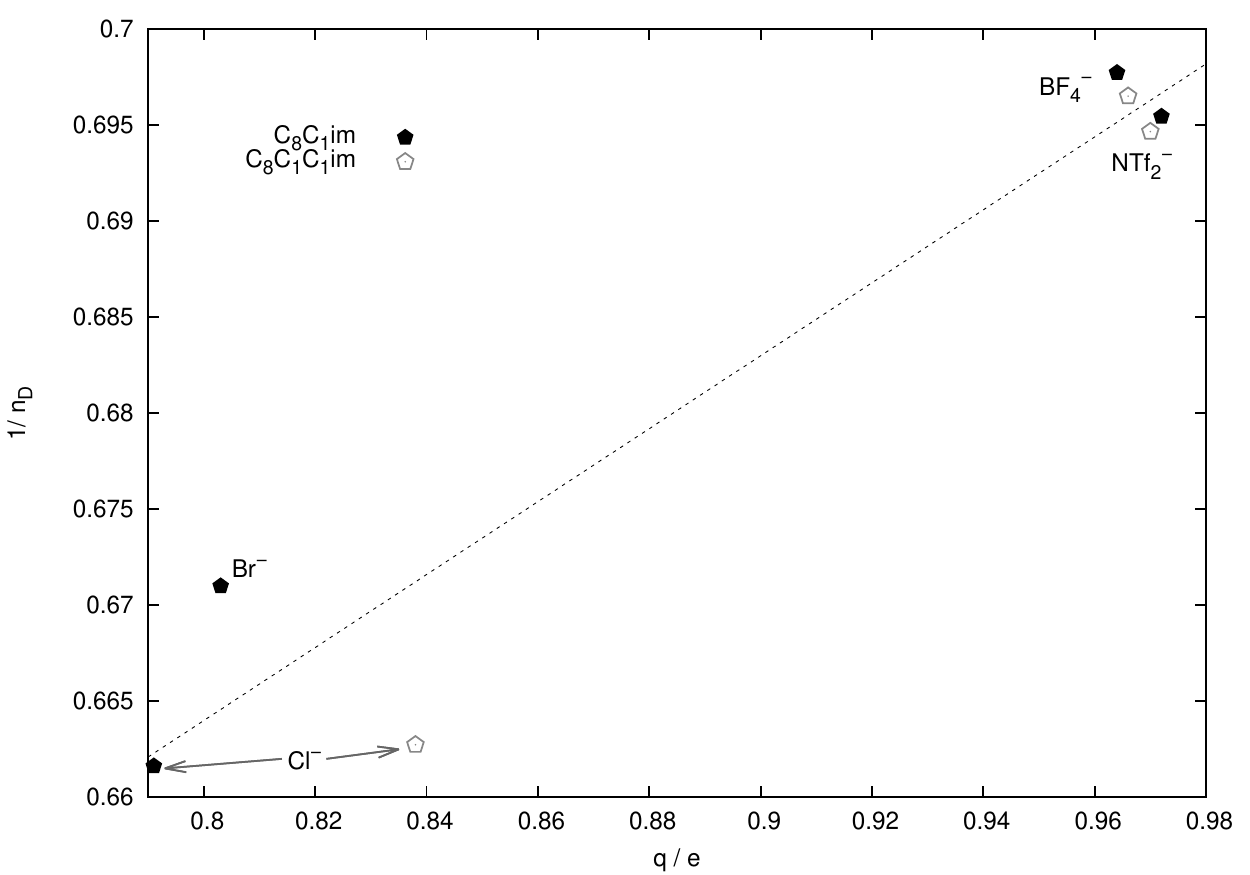}
 \caption{Inverse of the refractive index (calculated from our values in Table~\ref{TAB_DR}) of octylimidazolium based ionic liquids 
 as a function of the net charge of the cation $q^+$ derived from BLYP-D/TZVPP calculations of an ion pair in gas phase. The refractive index 
 of 1-octyl-3-methylimidazolium bromide $n_D = 1.49$ was taken from Ref.~\onlinecite{sha06b}.
 \label{FIG_q_c8mim}}
\end{figure}

The correlation between the fractional charges and the inverse of the refractive index also holds true for octylimidazolium-based ionic liquids of 
Ref.~\onlinecite{ste10c} as visible in Fig.~\ref{FIG_q_c8mim}
although a simple correlation between the molecular polarizability $\amol$ and the fractional molecular charges is not possible anymore.
This may be due to several reasons: 
In Ref.~\onlinecite{ste10c} the partial charges were derived from a natural population analysis of BLYP-D/TZVPP calculations which may lead to 
a different charge distribution compared to our ChelpG results using MP2/aug-cc-pVDZ.
Another reason may be the stronger coordinating properties of chloride compared to tetrafluoroborate or NTf$_2^-$ or its smaller volume.
The latter reason is supported by the fact, that the correlation between the fractional charges and the inverse of the $\nD(\amol,\Vmol)$
is found. However, the fact that in-plane or on-top configurations in Ref.~\onlinecite{ste10c} has only little impact on the charge scaling 
factor supports the argumentation of a collective non-specific interaction like charge delocalization rather than local specific charge 
transfer coupling.

\section{Conclusion}
The refractive index and the mass density of ionic liquids can be predicted by our model of additive atomic polarizabilities and atomic volumes,
in particular for imidazolium based ionic liquids, with reasonable accuracy. Moreover, the agreement between our predictions and quantum-chemical calculations 
on a MP2/aug-cc-pVDZ level is quite good for neutral molecules. However, our Designed Regression model overestimates the polarizabilities of cations and underestimates
corresponding values for the anions. This can be corrected by the model of the excess electron. 

The fractional value of the net charge of cations and anions in quantum-chemical gas phase calculations should be interpreted in terms of polarizabilities
since the scaling factor decreases with increasing chain length of alkylammonium tetrafluoroborates although the net charge of the polar head NH$_3^+$-group
oscillates around a constant value. Furthermore, the charge of NH$_3^+$ does not change very much if the anion has left.
In MD simulations these fractional charges represent to some extent the averaged effect of polarizable forces. 

The present work constitutes the necessary first step towards the development of a polarizable force field based on the CL\&P parameterisation.\cite{pad04a,pad12a} 
Future efforts (either by us or other groups) will be able to use the atomic polarizability values described herein to anchor and develop systematic and 
general models that encompass the possibility of local and responsive polarization of the molecular ions of ILs and also molecular solvents or substrates.
Here, the reparametrization of Lennard-Jones and/or Buckingham potentials is one of the next steps since the induced dipolar interaction modelled by the 
polarizable forces on the basis of the atomic polarizabilities is already accounting for dispersion.\cite{rou03a,rou10a}
Consequently, the attractive part of corresponding Lennard-Jones or Buckingham potentials has to be adjusted. 

\section{Acknowledgement}
C.~Schr\"oder acknowledges the support of his Short term mission grant from COST action CM1206. The researchers from Portuguese institutions were supported by 
Funda\c{c}\~{a}o para a Ci\^{e}ncia e Tecnologia (FCT) through projects FCT-ANR/CTM-NAN/0135/2012 and UID/QUI/00100/2013. 
K.~Shimizu acknowledges postdoctoral grant SFRH/BPD/94291/2013.


\end{document}